\documentclass[batteries,article,accept,pdftex,moreauthors]{Definitions/mdpi} 
\usepackage{tabularx}

\firstpage{1} 
\pubvolume{1}
\issuenum{1}
\articlenumber{0}
\pubyear{2023}
\copyrightyear{2023}
\externaleditor{Academic Editor: }
\datereceived{} 
\dateaccepted{} 
\datepublished{} 

\hreflink{https://doi.org/}

\Title{Off-resonant Dicke Quantum Battery: Charging by Virtual
Photons}

\TitleCitation{Off-resonant Dicke Quantum Battery: Charging by Virtual Photons}

\Author{
Giulia Gemme$^{1}$, Gian Marcello Andolina$^{2}$, Francesco Maria Dimitri
Pellegrino$^{3,4,5}$, Maura Sassetti$^{1,6,*}$ and Dario Ferraro$^{1,6,\orcidA{}}$
}

\AuthorNames{Giulia Gemme, Gian Marcello Andolina, Francesco Maria Dimitri
Pellegrino, Maura Sassetti and Dario Ferraro}

\AuthorCitation{Gemme, G.; Andolina, G.M.; Pellegrino, F.M.D.; Sassetti, M.; Ferraro, D.}

\address{%
$^{1}$ \quad Dipartimento di Fisica, Università di Genova, Via Dodecaneso 33, 16146 Genova, Italy; 
giulia.gemme@edu.unige.it (G.G.);  dario.ferraro@unige.it (D.F.)\\

$^{2}$ \quad JEIP, USR 3573 CNRS, Collège de France, PSL Research University, 11 Place Marcelin Berthelot, F-75321 Paris, France; 
gian-marcello.andolina@college-de-france.fr (G.M.A.)\\
$^{3}$ \quad Dipartimento di Fisica e Astronomia "Ettore Majorana", Università di Catania, Via S. Sofia 64, I-95123 Catania, Italy;
francesco.pellegrino@dfa.unict.it (F.M.D.P.) \\
$^{4}$ \quad INFN, Sez. Catania, I-95123 Catania, Italy; \\
$^{5}$ \quad CNR-IMM, Via S. Sofia 64, I-95123 Catania, Italy; \\
$^{6}$ \quad CNR-SPIN, Via Dodecaneso 33, 16146 Genova, Italy; \\
}

\corres{Correspondence: sassetti@fisica.unige.it}

\abstract{We investigate a Dicke quantum battery in the dispersive regime, where the photons trapped into a resonant cavity are way more energetic with respect to the two-level systems embedded into it. Under such off-resonant conditions, even an empty cavity can lead to the charging of the quantum battery through a proper modulation of the matter-radiation coupling. This counterintuitive behaviour has its roots in the effective interaction between two-level systems mediated by virtual photons emerging from the fluctuations of the quantum electromagnetic field. In order to properly characterize it, we address relevant figures of merit such as the stored energy, the time required to reach the maximum charging, and the averaged charging power. Moreover, the possibility of efficiently extracting energy in various ranges of parameters is discussed. The scaling of stored energy and power as a function of the number $N$ of two-level systems and for different values of the matter-radiation coupling is also discussed showing, in the strong coupling regime, performances in line with what reported for the Dicke quantum battery in the resonant regime.}


\begin{document}

\section{Introduction}

Quantum batteries (QBs)~\cite{Alicki13} ---genuinely non-classical devices capable of storing energy and performing useful work--- have been a very active topic of research in recent years~\cite{Campaioli18, Bhattacharjee21}. In this framework, theoretical studies have demonstrated that entangling operations can consistently accelerate the charging~\cite{Binder15}, with the Dicke QB standing out as a particularly promising candidate~\cite{Ferraro18, Crescente20, Dou22c, Dou22, Erdman22}. It is given by a system where the energy of photons trapped into a cavity (acting as the charger) is coherently transferred to a QB composed of $N$ quantum units, namely a collection of identical independent two-level systems (TLSs). This model describes, for example, an ensemble of real or artificial atoms embedded in a Fabry-Perot cavity, namely a resonator consisting of two parallel, highly reflective mirrors placed a certain distance apart~\cite{Dicke54}. 

In this system, the energy is initially stored into the cavity and flows towards the TLSs due to a proper modulation of the matter-radiation coupling~\cite{Andolina18}. In order to favour this energy flow the cavity and the TLSs are usually assumed as resonant, namely characterized by the same typical energy. Once the charging process is complete, the coupling between the two systems is turned off, leaving the TLSs in a charged state whose stability crucially depends on their relaxation and dephasing times~\cite{Farina19, Santos19, Barra19, Carrega20, Morrone22, Gemme22}. This kind of device displays a collective speed-up of the charging process. This fact has attracted a great deal of interest and, very recently, a first step towards the realization of a Dicke QB has been experimentally implemented in an excitonic system made of fluorescent organic molecules dispersed in an inert matrix, where the Dicke superabsorption leads to a collective boost of the charging process~\cite{Quach22}. Despite this remarkable achievement, the investigation of this kind of device is still in its infancy and new theoretical and experimental studies could lead to promising routes towards the exploitation of purely quantum effect in miniaturized devices devoted to energy storage. 

In this direction we want to investigate the role played by exchange of virtual photons between the TLSs composing the QB. A paradigmatic example of the dynamical effect of virtual photons is the Casimir force which arises from the quantum fluctuations of the electromagnetic field. It was first theoretically investigated by Hendrik Casimir in 1948~\cite{Casimir48} and experimentally proved almost fifty years later~\cite{Lamoureaux97}. This force is due to the presence of virtual photons in the vacuum and, in the case of a Fabry-Perot cavity, depends on the boundary conditions imposed by the mirrors. In particular, in the case of two parallel, uncharged, and perfect metallic plates, the virtual cavity photons are restricted in their fluctuation modes by the plates themselves. As a result, the plates experience a net attractive force. 

Motivated by this, we introduce and characterize a new type of device for energy storage where the charging is mediated by virtual photons. In contrast to the conventional Dicke QB, the cavity in our proposed off-resonant Dicke QB can be also empty and the two systems (charger and QB) are far detuned in energy, preventing any direct energy flow between them. This makes the systems easier to realize with respect to the conventional Dicke QB and allows us to describe it in the so-called dispersive regime~\cite{Schleich_Book}, typically used in quantum computing for qubit readout~\cite{Krantz19} and characterized by an effective infinite range interaction among the TLSs of the Lipkin-Meshkov-Glick (LMG) kind~\cite{Lipkin65, Dou22b}. Our charging protocol works as follows: initially, both the cavity and the QB are in their ground states and decoupled. The cavity mirrors are then brought close together, leading to a finite matter-radiation coupling. Thus, the virtual detuned photons excite the TLSs and charge the QB. The mirrors are then moved away, switching-off again the matter-radiation coupling leaving the atoms in an excited state. The use of virtual photons in this process gives the QB a unique {\it quantum} behaviour. Remarkably enough, in the strong coupling regime, the performances of these off-resonant Dicke quantum batteries are comparable with the ones of the resonant case.  

The article is organized as follows. In Section~\ref{Model} we consider the dispersive regime of the conventional Dicke QB with dipolar matter-radiation coupling. Under the assumption of highly energetic cavity photons, we derive an effective description of the off-resonant Dicke QB based on an LMG model characterized by an infinite range interaction among the TLSs and showing relevant simplifying conservation laws. The charging protocol based on switching on and off the matter-radiation coupling is also discussed. 
The main figures of merit, namely the energy stored into the QB, the averaged charging power, and the times required to reach their maxima are discussed in Section~\ref{Figures}. The ergotropy, namely the maximum energy which can be extracted by means of either collective or individual unitary operation is also introduced. Section~\ref{Results} reports the results concerning the time evolution of the above quantities for different values of the matter-radiation coupling and at various numbers $N$ of TLSs. The scaling of their maxima as a function of $N$ is also considered. A universal behaviour of the maximum stored energy is discussed in connection with the quantum phase transition predicted for the underlying effective model in Section~\ref{Universality}. Section~\ref{Conclusione} is devoted to the conclusions, while technical details of the calculations are reported in three Appendices.  

\section{Model}
\label{Model}

Let's start by considering a conventional Dicke-QB with a dipole matter-radiation coupling between $N$ identical TLSs and a resonant cavity~\cite{Ferraro18}. It is described by the Hamiltonian (from now on we will set $\hbar=1$)
\begin{eqnarray}
\hat{H}_{\rm {Dicke}}&=&\hat{H}_{B}+\hat{H}_{C}+\hat{H}_{B-C}\\
&=&\omega_z\hat{S}_z+\omega_c\hat{a}^{\dagger}\hat{a}+ 2\lambda\hat{S}_x\left(\hat{a}^{\dagger}+\hat{a}\right),
\label{Dicke}
\end{eqnarray}
with $\hat{H}_{B}$ the Hamiltonian of the QB, $\hat{H}_{C}$ the one of the cavity and $\hat{H}_{B-C}$ their coupling. In the above equation, we have introduced the collective pseudo-spin operator 
\begin{equation}
\hat{S}_{\alpha}=\frac{1}{2}\sum^{N}_{i=1}\hat{\sigma}^{i}_{\alpha}, 
\end{equation}
with $\hat{\sigma}^{i}_{\alpha}$ ($\alpha=x,y,z$) the conventional Pauli matrices associated with the $i$-th TLS and where $\hat{a}$ ($\hat{a}^{\dagger}$) are the annihilation (creation) operator for the photons in the cavity. Here, $\omega_{z}$ is the level spacing of each TLS, $\omega_{c}$ is the energy of the photons, and $\lambda$ is the intensity of the matter-radiation coupling. Notice that this configuration can be effectively engineered in conventional cavity~\cite{Walther06} and circuit-quantum electrodynamics~\cite{Blais04} set-ups. In the latter case, the TLSs can be realized by means of superconducting qubits coupled to an LC circuit playing the role of a resonant cavity~\cite{Krantz19}. 

According to the analysis carried out in Appendix~\ref{AppA}, considering the Schrieffer–Wolff transformation~\cite{Schleich_Book} up to the second order in $\lambda$ one obtains the effective Hamiltonian
\begin{equation}
    \hat{H}_{\rm eff}=\omega_z \hat{S}_z+\omega_c \hat{a}^{\dagger}\hat{a}-\frac{2\lambda^2\omega_z}{\omega_c^2-\omega_z^2}\hat{S}_z\left(\hat{a}^{\dagger}+\hat{a}\right)^2-\frac{4\lambda^2\omega_c}{\omega_c^2-\omega_z^2}\hat{S}_x^2,
\end{equation}
which is a good approximation of the Dicke Hamiltonian in Eq.~(\ref{Dicke}) in the dispersive regime $\lambda\ll|\omega_{c}-\omega_{z}|$~\cite{Schleich_Book}. Notice that, in this regime, the possibility to increase the energy stored into a QB through virtual excitations mediated by a higher excited state has been recently discussed~\cite{Santos23}. 

Further considering the condition of highly energetic photons $\omega_{c}\gg \omega_{z}$ one finally obtains~\cite{Roche22}
\begin{equation}
    \hat{H}'_{\rm eff}=\omega_z \hat{S}_z+\omega_c \hat{a}^{\dagger}\hat{a}-\frac{4\lambda^2}{\omega_c}\hat{S}_x^2.
    \label{H_1ph_primed}
\end{equation}
Getting rid of the cavity contribution it is possible to map Eq.~(\ref{H_1ph_primed}) into 
\begin{eqnarray}
\hat{H}&=&\hat{H}_{B}+\hat{H}_{\rm int}\\
&=&\omega_{z} \hat{S}_z-g\hat{S}_x^2,
\label{H}
\end{eqnarray}
with
\begin{equation}
g=\frac{4\lambda^2}{\omega_c}. 
\end{equation}
Notice that, according to this mapping, only positive values of the parameter $g$ are physically meaningful. Moreover, as shown in Appendix~\ref{AppA}, even a more exotic two-photon matter-radiation coupling~\cite{Felicetti15, Felicetti18} leads to an analogous effective Hamiltonian.

The Hamiltonian in Eq.~(\ref{H}) is characterized by an effective infinite range interaction between the TLSs of the LMG kind~\cite{Lipkin65, Dou22b, Abah22} mediated by virtual photons playing the role of a quantum bus~\cite{Larson10}. Such kind of infinite range interaction has been indicated as an ideal condition to achieve high power QBs~\cite{Campaioli17, Le18, Gyhm22}.

In the following, we will assume a charging protocol addressing a time-dependent version of Eq.~(\ref{H})
\begin{equation}
 \hat{H}(t) = \hat{H}_{B} + f(t)  \hat{H}_{\rm int} ~,
 \label{H_t}
\end{equation}
where $f(t)$ is a classical external control parameter introduced to turn on and off the effective interaction among the TLSs in Eq.~(\ref{H}). This can be ultimately seen as a direct consequence of the modulation of the matter-radiation coupling in the original Dicke model in Eq.~(\ref{Dicke}) obtained by bringing the cavity mirrors close together. For sake of simplicity, we will assume this control function to be step-wise: equal to $1$ for $t\in[0,\tau_{c}]$, with $\tau_{c}$ a given time which will be specified in the following, and zero elsewhere. It is worth noting that smoother charging protocols are expected to lead only to marginal changes in the behaviour discussed in the following~\cite{Crescente22}. Generally speaking, the possibility to optimize the time evolution of the charging protocol has been recently object of intense research for both the Dicke QB~\cite{Erdman22} and other QB technologies~\cite{Hu21, Rodriguez23}. Moreover, we further underline the fact that differently from what is usually considered~\cite{Campaioli18,Bhattacharjee21}, and in the same spirit of Refs.~\cite{Rosa20, Rossini20}, here the charger is encoded in the effective interaction among the TLSs composing the QB and not into some external classical or quantum device. 

\subsection{Conserved quantities}

In order to investigate the behaviour of QB, for a fixed number $N$ of TLSs, a full numerical exact diagonalization is needed. In order to consistently simplify this task some symmetries can be exploited. 

First of all, the operator 
\begin{equation}
\hat{S}^2=\hat{S}^2_x+\hat{S}^2_y+\hat{S}^2_z
\end{equation}
is a conserved quantity, namely $\left[\hat{H}(t),\hat{S}^2\right]=0$. This allows us to work in the $(N+1)\times (N+1)$ Hilbert space with maximal value $s=N/2$ and to characterize the states of the system in terms of a pseudo-spin basis $|s,m\rangle$ such that:
\begin{eqnarray}
    \hat{S}^2|s,m\rangle&=&s(s+1)|s,m\rangle,\\
    \hat{S}_z|s,m\rangle&=&m|s,m\rangle,
\end{eqnarray}
 with $m=-s, -s+1, \dots, s-1, s$.

Moreover, also the spin number parity operator
\begin{equation}
\hat{\Pi}=e^{-i\pi\hat{S}_z}
\end{equation}
satisfies $\left[\hat{H},\hat{\Pi}\right]=0$, leading to a further reduction of the Hilbert space dimension for proper initial conditions of the QB. According to this, in the following, we will investigate the charging of the QB starting from the ground state of $\hat{H}_{B}$
\begin{equation}
|\psi(0)\rangle=|\frac{N}{2}, -\frac{N}{2}\rangle.
\end{equation}
It is worth noticing that, according to this parity constraint, the perfect charging of the QB, namely the complete transition to the maximally charged state 
\begin{equation}
|\psi_{max}\rangle=|\frac{N}{2}, +\frac{N}{2}\rangle
\end{equation}
can only occur for an even number $N$ of TLSs. However, as will be clear in the following, the efficiency of this kind of QB is lower than $100\%$, therefore we can investigate equivalently both even and odd numbers of TLSs. 

\section{Figures of merit}
\label{Figures}

We introduce now relevant quantities able to characterize the performances of the considered off-resonant Dicke QB (with $\omega_{c}\gg \omega_{z}$). In the following, we will focus on the energy stored into the QB itself, the time required to reach the optimal charging, the energy trapped into the interaction term, the averaged charging power, and the ergotropy, namely the maximum amount of work extractable from a charged QB using either collective or local unitary operations~\cite{Alicki13, Allahverdyan04}.

\subsection{Stored energy and charging time}

First of all, we consider the energy stored into the QB. At a given time $t$ it is given by~\cite{Ferraro18, Andolina18}
\begin{equation}
E_{B}(t)=\langle \psi(t) |\hat{H}_{B}| \psi(t) \rangle-\langle \psi(0) |\hat{H}_{B}| \psi(0) \rangle,
\label{E_B}
\end{equation}
with $|\psi(t) \rangle$
the quantum state of the systems for $t\leq \tau_{c}$, evolved according to $\hat{H}(t)$.

The charging time, namely the time $t_{E}$ at which the quantity in Eq.~(\ref{E_B}) reaches is first maximum $E_{max}$, will be also investigated in the following. Of particular relevance will be the scaling of both
$E_{max}$ and $t_{E}$ as a function of the number $N$ of TLSs composing the QB for different values of the effective coupling $g$.

It is worth underlining the fact that the system under investigation is considered as closed. Under this approximation, which holds as long as the time evolution is shorter than the typical time scale associated with relaxation and dephasing of the TLSs~\cite{Dou22c, Carrega20}, one expects a net energy is injected into the QB via the modulation of the external control $f(t)$. 

\subsection{Averaged charging power}

Another relevant figure of merit is the averaged charging power, namely the energy stored in a given time, defined as~\cite{Binder15, Campaioli17}
\begin{equation}
P(t)=\frac{E_{B}(t)}{t}. 
\end{equation}
In analogy with what was done above, we are interested in the scaling of the maximum charging power $P_{max}$ and of the time $t_{P}$ requited to reach it as a function of $N$ and for various values of the effective coupling $g$. 

\subsection{Ergotropy}

Here we introduce the ergotropy, namely the maximum work extractable from a QB~\cite{Allahverdyan04}. We will focus both on the total ergotropy and the one associated with only one of the TLS composing the QB. On a very general ground at a given time $t$, for an Hamiltonian $\hat{\mathcal H}$ and a time dependent density matrix $\hat{\rho}(t)$ describing a quantum state at a given time, it is possible to define the quantity  

\begin{equation}\label{eq:ergotropy2}
\mathcal{E}\left(\rho(t),\hat{\cal H}\right)={\rm tr}[\hat{\mathcal{H}} \hat{\rho}(t)]-{\rm tr}[\hat{\mathcal{H}}  \hat{\pi}_{\rho}(t)]\;,
\end{equation}
with $\hat{\pi}_{\rho}(t)$ the passive state associated to $\hat{\rho}(t)$ at a given time. Given the diagonal representation of the density matrix 
\begin{equation}
\hat{\rho}(t)=\sum_n r_n(t) |r_n(t)\rangle\langle r_n(t)|,
\end{equation}
with  $r_0(t)\geq r_1(t) \geq \cdots$, the associated passive state is diagonal on the eigenbasis of $\hat{\mathcal{H}}$ ($|\epsilon_n\rangle$) and reads  
\begin{equation}
\hat{\pi}_{\rho}(t)=\sum_n r_n(t) |\epsilon_n\rangle\langle\epsilon_n|, 
\end{equation}
with energy eigestates ordered in such a way that $\epsilon_0\leq \epsilon_1 \leq \cdots$. This finally leads to the general definition

\begin{equation}
\mathcal{E}\left(\hat{\rho}(t),\hat{\cal H}\right)=\sum_{k,n} r_{k}(t)\epsilon_{n}\left(|\langle r_{k}(t)|\epsilon_{n}\rangle|^{2}-\delta_{k,n} \right), 
\end{equation}
which can be directly applied to our particular case considering 
\begin{equation}
\hat{\rho}(t)=|\psi(t)\rangle \langle \psi(t)|    
\end{equation}
and $\hat{\mathcal{H}}=\hat{H}_{B}$.

Due to the fact that we are considering only unitary operations acting on a closed system, the ergotropy defined above coincides with the energy stored into the QB~\cite{Alicki13}.

For what it concerns the ergotropy of a single TLS of the QB in~Eq.(\ref{Dicke}) one can proceed as follows. The $2\times 2$ reduced density matrix describing the first TLS at a given instant is indicated by $\hat{\rho}_{{B},1}$, while the energy is measured with respect to the Hamiltonian 
\begin{equation}
\hat{h}^1_{B}=\frac{\omega_{z}}{2}\hat{\sigma}^{1}_{z}.
\end{equation}
Notice that, due to the invariance under TLSs permutations of the Hamiltonian in Eq.~(\ref{H}), we can focus on the first TLS without any loss of generality. Thus, the maximum energy that can be extracted from a single battery unit in an $N$ TLSs device is given by 
\begin{eqnarray}
\mathcal{E}_{1}^{(N)}(t) \equiv \mathcal{E}(\hat{\rho}_{{B},1}(t),\hat{h}^1_{B})~. \label{ERGO}
\end{eqnarray}
 Consistently with what is done above, this expression can be further simplified by writing $\hat{\rho}_{{B},1}(t)$ in the diagonal basis,
\begin{eqnarray}
\hat{\rho}_{{B},1}(t)= r_{0}(t)|r_0(t)\rangle\langle r_0(t)|+r_{1}(t)|r_1(t)\rangle\langle r_1(t)|~, \label{rho_diag}
\end{eqnarray}
where the eigenvalues are again ordered in such a way that $r_0(t)\geq r_1(t)$. In this case, the ergotropy $\mathcal{E}_{1}^{(N)}(t)$ reduces to
\begin{eqnarray}
\mathcal{E}_{1}^{(N)}(t) =\frac{E^{(N)}(t)}{N}-r_{1}(t)\omega_{z}~, \label{ERGOS}
\end{eqnarray}
where we used the fact that ${\rm tr}[\hat{\rho}_{{B},1}(t)\hat{h}_1^{B}]=({E^{(N)}(t)}/{N})$, with $E^{(N)}={\rm tr}[\hat{H}_{B} \hat{\rho}(t)]$, due to permutation symmetry. We have evaluated the above expression numerically, further details about this analysis can be found in Ref.~\cite{Erdman22}.

We conclude this Section by pointing out that for a fixed number $N$ of TLSs, in general,
\begin{equation}
\mathcal{E}(t)\neq N\mathcal{E}^{(N)}_{1}(t).
\end{equation}
This is a consequence of the fact that local unitary operations acting on a unique TLS are not enough in order to extract all the energy stored into the QB, and non-local unitary operations collectively acting on the whole QB are needed in order to extract also the energy trapped in the correlations among the elementary building blocks composing the device~\cite{Alicki13}.

\section{Results and scaling laws}
\label{Results}

In the following, we will investigate in detail the behaviour of the previously discussed figures of merit for different values of the coupling $g$ and as a function of the number $N$ of TLSs composing the off-resonant Dicke QB. 

\subsection{Weak coupling}
\begin{figure}[p]
    \centering
    \includegraphics[width=\textwidth]{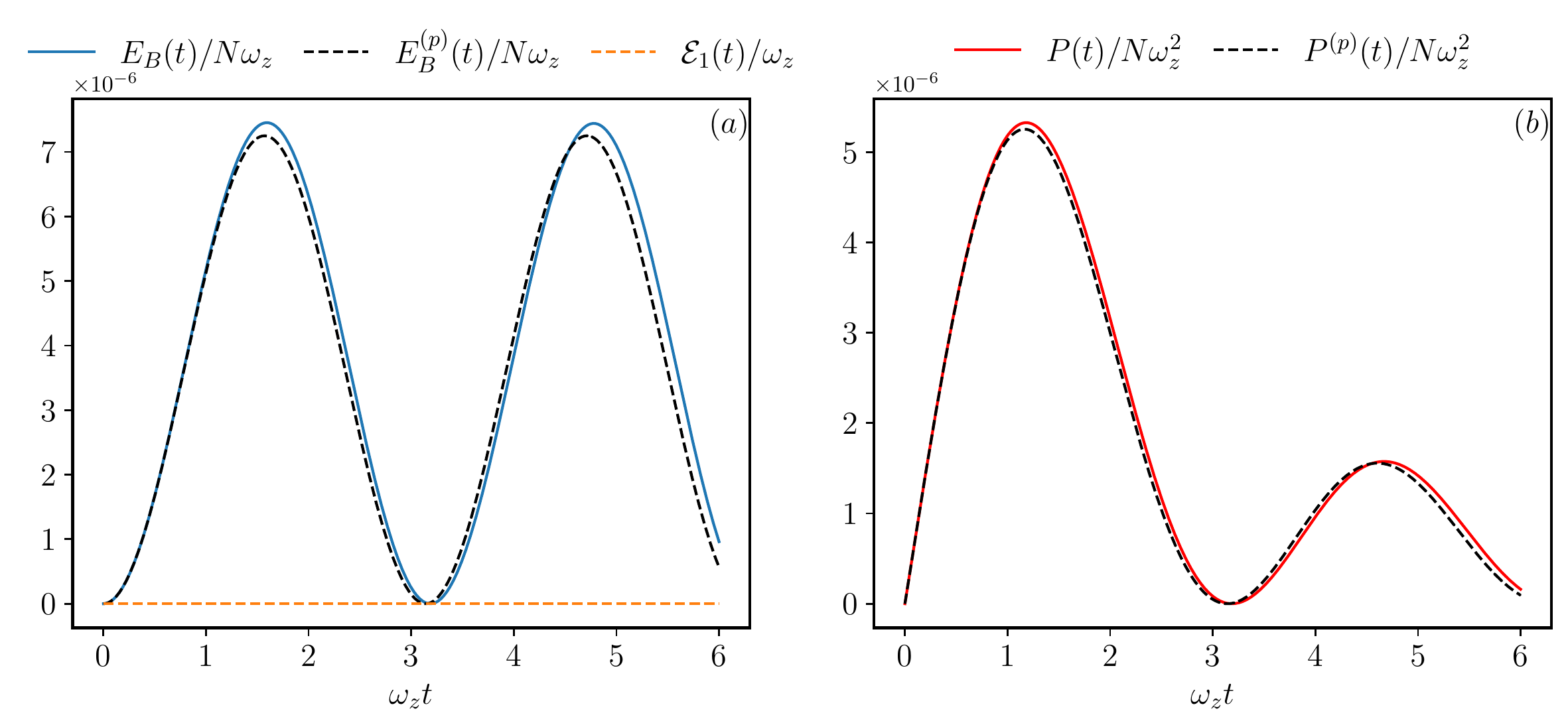}
     \caption{(a). Energy stored into the QB $E_{B}(t)$ (in units of $N\omega_{z}$) and single TLS ergotropy $\mathcal{E}_{1}(t)$ (in units of $\omega_{z}$) as a function of time (in units of $\omega^{-1}_{z}$) for $N=30$. (b). Averaged charging power $P(t)$ (in units of $N\omega^{2}_{z}$) as a function of time (in units of $\omega^{-1}_{z}$) for $N=30$. Other parameters are  $g=10^{-3}\omega_z$ and $\omega_{z}\tau_{c}>2\pi$. Black curves represent the theoretical curves obtained through time dependent perturbative expansions (indicated with the superscript $p$ in the labels) in Eqs.~(\ref{E_small}) and (\ref{P_small}).}
    \label{Fig1}
\end{figure}

Let's start by considering the regime $g\ll\omega_{z}$. The time evolution of the energy stored in the QB and the averaged charging power are reported in Fig.~\ref{Fig1}. 

 This behavior can be understood considering the analytical expressions for both the above quantities using a time dependent perturbative approach (see Appendix \ref{AppB} for more details about the derivation). This leads to 
\begin{equation}
    E_{B}(t)\approx\frac{g^2}{8}\frac{1-\cos{2\omega_z t}}{\omega_z}\left(N^2-N\right),
    \label{E_small}
\end{equation}
and
\begin{equation}
    P(t)\approx\frac{g^2}{8}\frac{1-\cos{2\omega_z t}}{\omega_z t}\left(N^2-N\right).
    \label{P_small}
\end{equation}

These approximated expressions correspond to the black dashed curves reported in Fig.~\ref{Fig1}. For the reported coupling $g=10^{-3}\omega_{z}$ they show an excellent agreement with the numerical curves up to $N\approx 30$, further validating the correctness of the considered exact diagonalization  and supporting the observation that there is only one relevant frequency of the state dynamics in this regime, leading to very smooth curves with no beats. It is worth noting that the amount of energy stored in this regime is extremely small (see the scale in Fig.~\ref{Fig1}). As a consequence of this, the ergotropy for a single TLS is zero (dashed orange curve in Fig.~\ref{Fig1}), namely no energy can be extracted for a single TLS using local unitary operations.  

\begin{figure}[p]
         \centering
         \includegraphics[width=\textwidth]{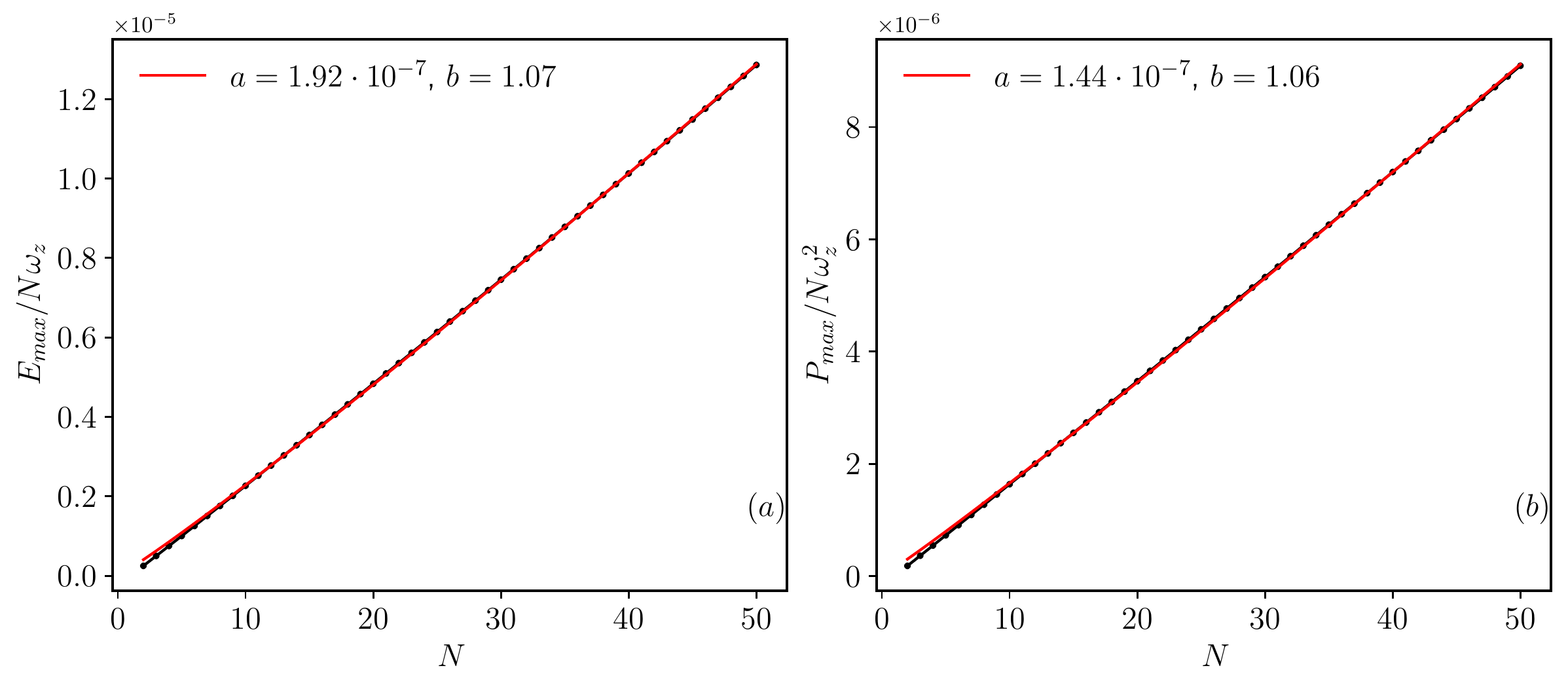}
    \caption{(a). Maximum stored energy $E_{max}$ (in units of $N\omega_{z}$) and (b) maximum averaged charging power $P_{\max}$ (in units of $N\omega_{z}$) as a function of the number of TLSs $N$. Other parameters are $g=10^{-3}\omega_z$, $\tau_{c}=t_{E}$ in (a) and $\tau_{c}=t_{P}$ in (b). Red curves are the fits of the numerical points according to: $E_{max}/N\omega_z=a N^{b}$ and $P_{max}/N\omega^{2}_z=aN^{b}$. The values of the fitting parameters are indicated in the labels of each panel.}
    \label{Fig2}
\end{figure}

Starting from Eq.~(\ref{E_small}) one can obtain the values of the maximal stored energy and of the charging time associated with the first maximum, namely
\begin{eqnarray}
    E_{max}&=&\frac{g^2}{4\omega_z}\left(N^2-N\right),\\
    \omega_{z} t_{E}&=&\frac{\pi}{2},
\end{eqnarray}
respectively. Analogously for the averaged charging power in Eq.~(\ref{P_small}) one has
\begin{eqnarray}
    P_{max}&\approx&\frac{g^2}{4.46}\left(N^2-N\right),\\
    \omega_{z}t_{P}&\approx& 1.16,
\end{eqnarray}
where the values in this case need to be extracted numerically. The validity of the scalings of stored energy and charging power as a function of the number of TLSs has been checked in Fig.~\ref{Fig2}. In this framework, we underline the fact that throughout all the paper we will consider the scaling at a finite number of TLSs, without addressing the thermodynamical limit, differently from what is discussed for example in Ref.~\cite{Julia-Farre20}.  This is justified by the fact that, for example, circuit quantum-electrodynamics devices are usually far from the thermodynamic limit~\cite{Fink09}.

\subsection{Strong coupling}
\begin{figure}[p]
    \centering
    \includegraphics[width=\textwidth]{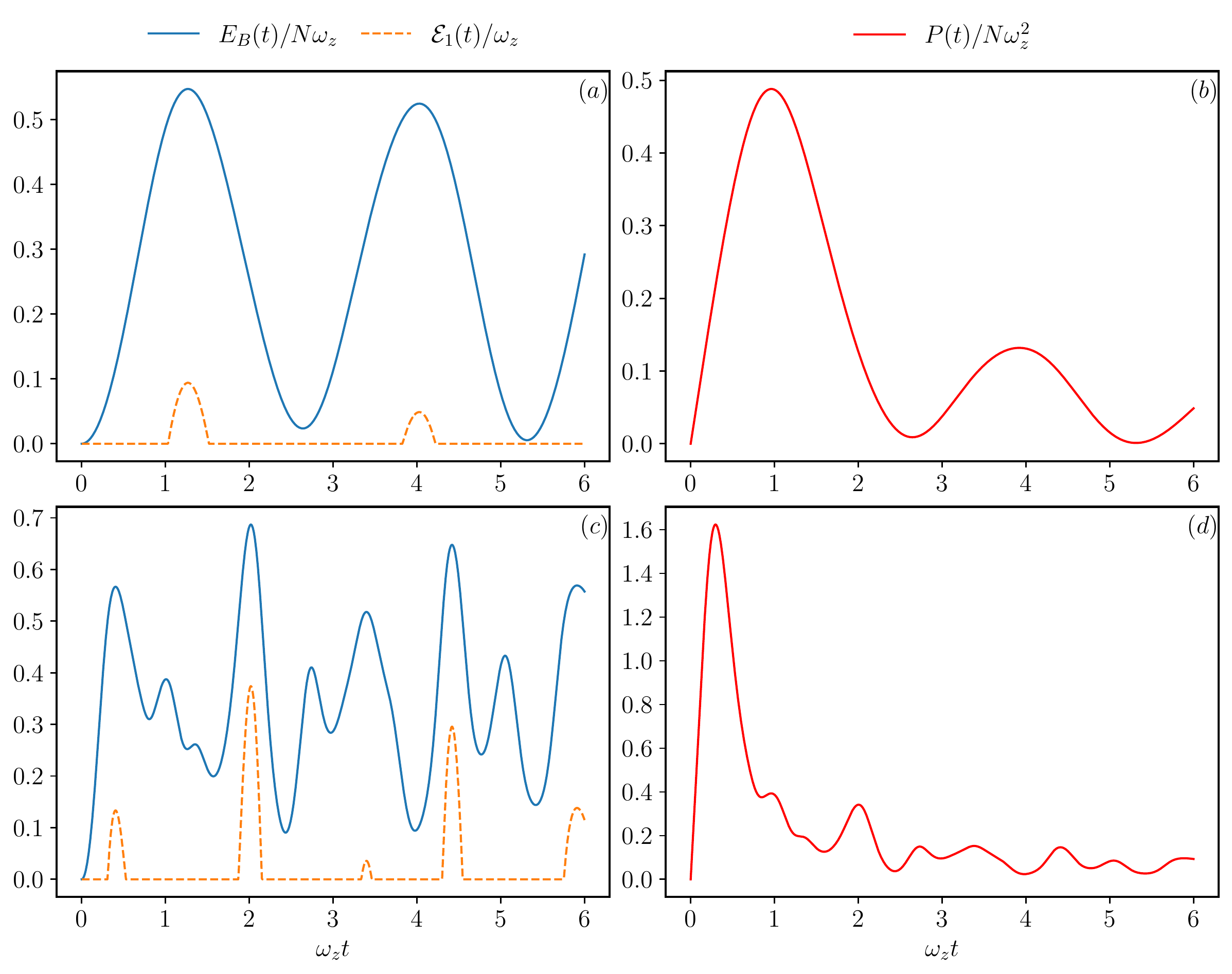}
    \caption{Energy stored into the QB $E_{B}(t)$ (in units of $N\omega_{z}$) and single TLS ergotropy $\mathcal{E}_{1}(t)$ (in units of $\omega_{z}$) as a function of time (in units of $\omega^{-1}_{z}$) for $N=4$ (a) and $N=30$ (c). Averaged charging power $P(t)$ (in units of $N\omega^{2}_{z}$) and a function of time (in units of $\omega^{-1}_{z}$) for $N=4$ (b) and $N=30$ (d). Other parameters are $g=\omega_z$ and $\omega_{z}\tau_{c}>2\pi$.}
    \label{Fig3}
\end{figure}

We investigate here the strong coupling regime, namely a situation in which $g$ is of the same order of $\omega_{z}$ \cite{Giannelli22}. Even if difficult to be realized in the dissipative Dicke model with dipolar coupling, this situation could be reached by exploiting more exotic matter-radiation couplings such as the two-photon interaction and properly changing the initial photon distribution in the cavity (see Appendix \ref{AppA} for more details). The time evolution of the figures of merits, reported in Fig.~\ref{Fig3}, appears again quite regular at small $N$ but shows a complicated beating structure at greater values of $N$. Notice that, in this regime, the first maximum of the energy stored into the QB is only a local maximum, with higher values of the energy possibly appearing at greater times. However, we focus on it due to the fact that it is characterized by a greater value of the power (see Fig.~\ref{Fig3} (d)). In this regime, the charging of the QB can exceed the $50\%$ near the maxima. In correspondence to these points, one observes the emergence of a non-zero single TLS ergotropy (see the dashed orange curve in Fig.~\ref{Fig3} (a) and (b)), the signature of the possibility to extract energy from the QB also by using local unitary operations. 

\begin{figure}[p]
     \centering
         \includegraphics[width=\textwidth]{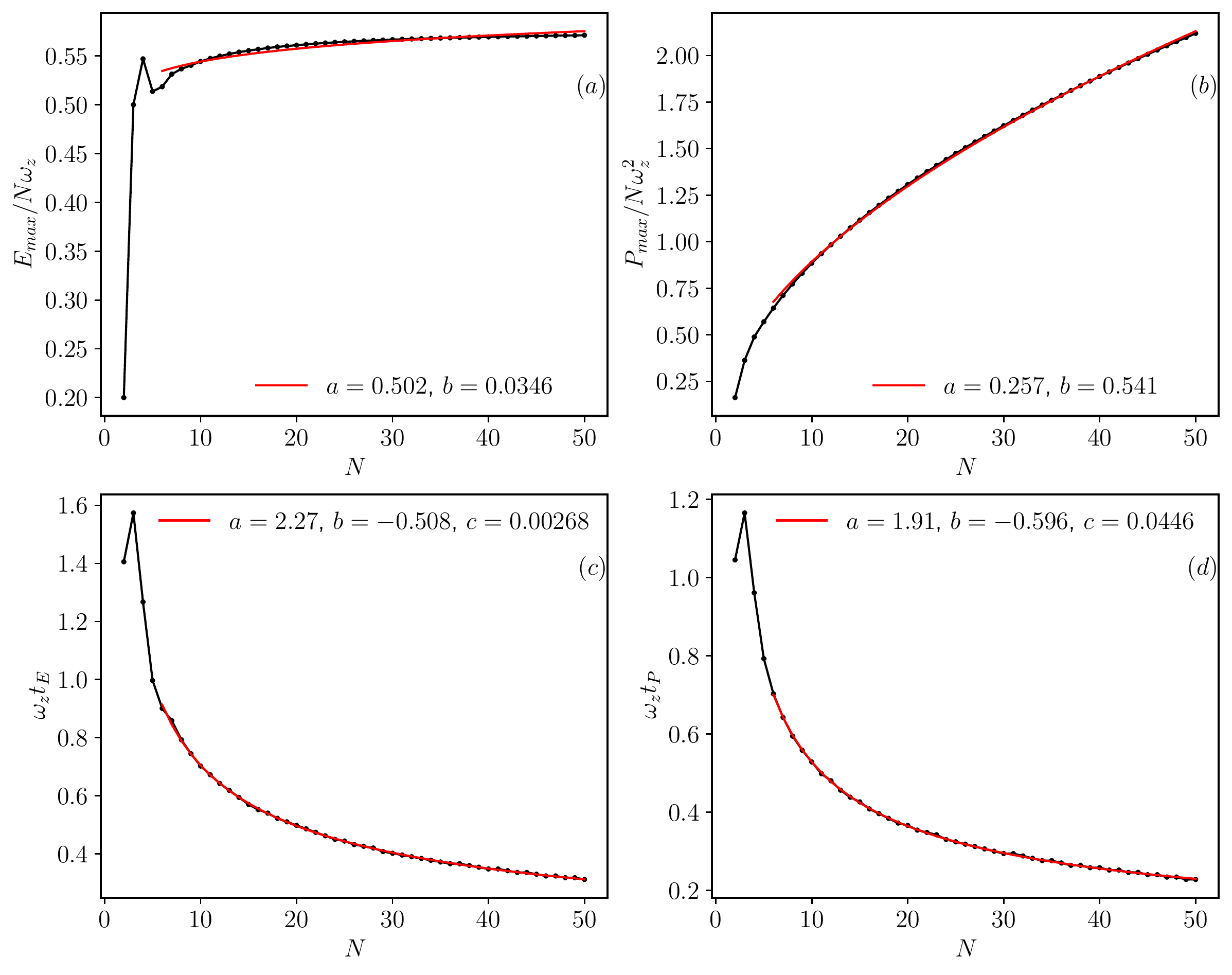}
    \caption{Maximum stored energy $E_{max}$ (in units of $N\omega_{z}$) (a) and corresponding charging time $\omega_{z}t_{E}$ (c) as a function of the number of TLSs $N$. Maximum averaged charging power $P_{\max}$ (in units of $N\omega^{2}_{z}$) (b) and corresponding charging time $\omega_{z}t_{P}$ (d) as a function of the number of TLSs $N$. Other parameters are $g=\omega_z$, $\tau_{c}=t_{E}$ for each $N$ in (a,c) and $\tau_{c}=t_{P}$ for each $N$ in (b,d). Red curves are the fits of the numerical points according to: $E_{max}/N\omega_z=aN^{b}$ (a), 
    $P_{max}/N\omega^{2}_z=aN^{b}$ (b), $\omega_{z}t_{E}=aN^{b}+c$ (c),  and $\omega_{z}t_{P}=aN^{b}+c$ (d). The values of the fitting parameters are indicated in the labels of each panel.}
    \label{Fig4}
\end{figure}

According to the full numerical analysis reported in Fig.~\ref{Fig4} it is possible to deduce the scaling behaviour of the various figures of merit in the strong coupling regime, namely

\begin{eqnarray}
E_{max}&\propto& N,\\
\omega_{z}t_{E}&\propto&N^{-\frac{1}{2}},\\
P_{max}&\propto& N^{\frac{3}{2}},\\
\omega_{z}t_{P}&\propto&N^{-\frac{1}{2}}.
\end{eqnarray}
Notice that these are reminiscent of the ones reported for the Dicke model with conventional dipole matter-radiation coupling~\cite{Ferraro18}. This can be justified by considering the fact that in this case the expectation value of the operator $\hat{S}_{x}$ over both the ground and the excited states of the system becomes extensively large, similarly to what happens for both $\hat{S}_{x}$ and the operator $(\hat{a}^{\dagger}+\hat{a})$ in the original Dicke model in Eq.~(\ref{Dicke}) after the superradiant quantum phase transition~\cite{Emary03}.

\section{Considerations about universality}
\label{Universality}

The previous analysis showed two drastically different scaling behaviours of the considered quantities for different regimes of the coupling, with the weak coupling showing a better scaling with $N$ and the strong coupling regime more promising in terms of absolute values of the energy storage, charging power, and ergotropy.  These are summarized in Table~\ref{Table1}. Notice that for intermediate values of the coupling, a crossover between the two behaviour emerges (not shown).

\begin{table}[h!]
  \begin{center}
    \begin{tabular}{c|c|c|c} 
      Coupling regime & $E_{max}$ & $P_{max}$ & $t_{E}, t_{P}$\\
      \hline
      Weak coupling & $N^{2}$ & $N^{2}$ & Constant\\
      Strong coupling & $N$ & $N^{\frac{3}{2}}$& $N^{-\frac{1}{2}}$\\
    \end{tabular}
  \end{center}
    \caption{Scaling behaviour for some figures of merit in the weak and strong coupling regimes.}
        \label{Table1}
\end{table}

\begin{figure}[h]
     \centering
     \includegraphics[width=\textwidth]{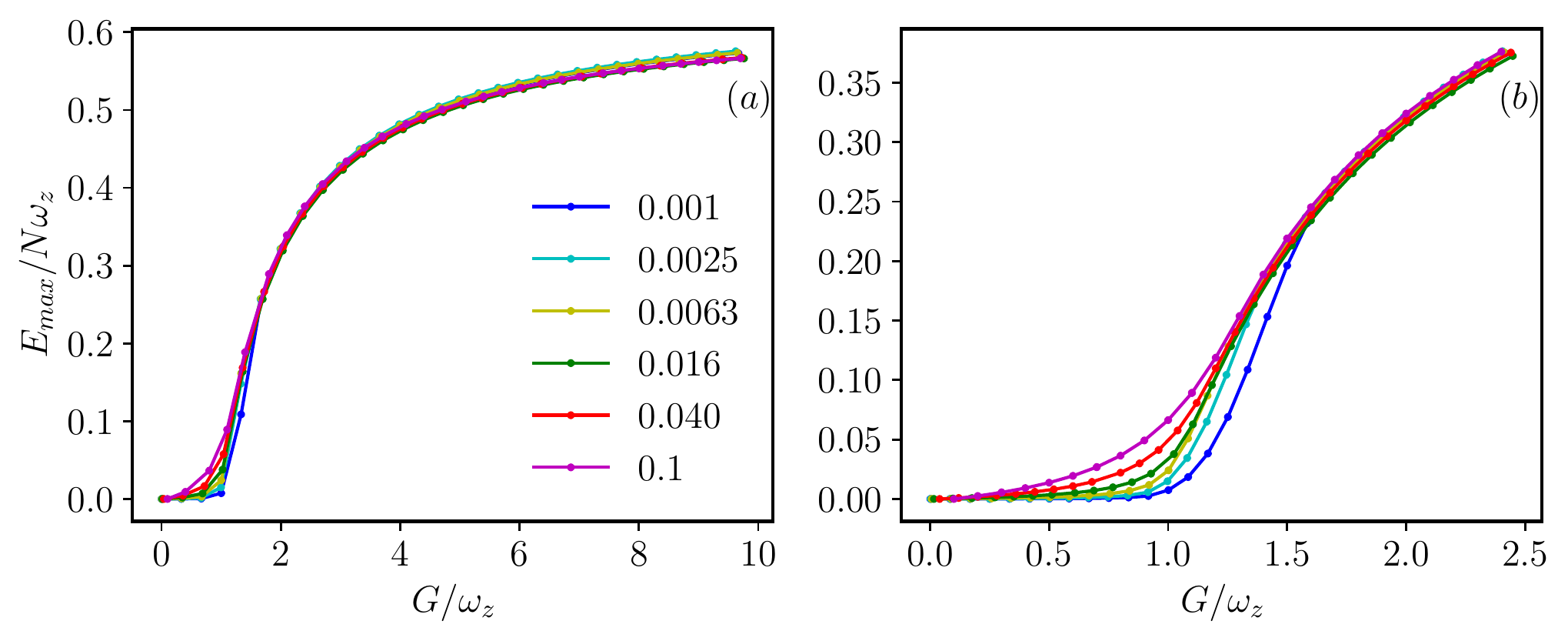}
    \caption{Behaviour of $E_{max}$ (in units of $N\omega_{z}$) as a function of the rescaled coupling $G=gN$ (in units of $\omega_{z}$) for various values of $g$ (a) and its zoom around the crossover at $G=\omega_z$ (b). Values of the coupling are reported in the color scale of the legend, while we have assumed $\tau_{c}=t_{E}$ for each value of $N$.}
    \label{Fig5}
\end{figure}

This change of behaviours can be seen as a consequence of the Excited State Quantum Phase Transition (ESQPT) shown by both the Dicke and the effective LMG model (see Ref.~\cite{Cejnar21} for a review on the subject). This represents the footprint of a quantum phase transition at the level of the excited states of the system. Its role in the present QB is justified by the fact that it is initialized in the ground state and charged by means of a protocol which naturally involves the excited states. In order to better clarify this point, we can consider the evolution of $E_{max}$ as a function of the renormalized coupling $G=gN$ (see Fig.~\ref{Fig5}). The choice of this renormalized parameter to characterize the evolution of the system can be justified by considering the classical limit of the effective LMG model (see Appendix~\ref{AppC} for more details) and reminiscent of the renormalization $\Lambda=\lambda \sqrt{N}$ observed for the Dicke model~\cite{Ferraro18}. Notice that the emergence of a crossover behaviour for the maximum stored energy is motivated by the fact that it is closely related to the order parameter of the system, namely the total magnetization, which at a given time $t$ can be written as 
\begin{equation}
\mathcal{M}(t)=\langle \psi(t) | \hat{S}_{z} |\psi(t)\rangle= \frac{E_{B}(t)}{\omega_{z}}+\langle \psi(0) | \hat{S}_{z} |\psi(0)\rangle.
\end{equation}

It is possible to observe now two different regimes for $G<\omega_z$ and $G>\omega_z$ respectively, with a crossover in correspondence of the ESQPT of the model, occurring at $G=\omega_z$. Incidentally, by increasing the value of $G$ (Fig.~\ref{Fig5} (b)) one has that the maximum value of energy per TLS reachable in the considered QB can exceed, the $60 \%$. Remarkably enough, this value is compatible with what is observed for the Dicke model in the strong coupling regime and under resonant conditions~\cite{Ferraro18}. This makes the present off-resonant Dicke QB competitive for practical experimental realizations, also taking into account the fact that it can be realized also using a vacuum cavity which is easier to be engineered and more stable with respect to what happens in the resonant Dicke QB. Moreover, the dependence on the renormalized coupling $G$ allows to extend the validity of the previous analysis carried out for the strong coupling regime by increasing the number of TLS composing the QB.

\section{Conclusions}
\label{Conclusione}

We have investigated an off-resonant Dicke QB, namely a device composed by $N$ TLSs embedded into a highly detuned resonant cavity. Under this condition, the TLSs feel an effective infinite range interaction among them. By properly switching on and off this coupling it is possible to promote at least a part of the TLSs from the ground to the excited states, thus charging the QB, even in the case of an empty cavity. This phenomenology can be ultimately related to the exchange of virtual photons whose coupling with the TLSs is varied bringing the mirrors of the cavity close together. 

By looking at relevant figures of merits such as the stored energy, the averaged charging power, and the time required to reach their respective maxima, we have characterized the performance of this kind of system. We have identified a small coupling regime, showing a very poor charging and no possibility of energy extraction (single TLS ergotropy) despite the relevant scaling in both the energy and the averaged charging power, and a strong coupling limit showing the same collective advantage in averaged charging power of the conventional Dicke QB~\cite{Ferraro18} with dipole matter-radiation interaction and in resonant conditions. The crossover between these two regimes is related to the quantum phase transition occurring in the effective model describing the QB~\cite{Cejnar21}. 

The present device shows, above the critical coupling, performances comparable to the ones of the Dicke QB in the resonant regime. In addition, the observed physics can be obtained also working with an empty cavity, thus avoiding the need of controlling the initial quantum state of radiation, the physically relevant parameters of the system, or exotic matter-radiation interactions~\cite{Crescente20}. These remarkable properties make it a very convincing candidate for future practical implementations of a functioning QB in various platforms  already considered in the framework of the Dicke QB, such as superconducting qubits or array of semiconducting quantum dots coupled to an LC circuit through a tunable capacitance~\cite{Xiang13, Stockklauser17}.

\vspace{6pt}
\authorcontributions{Conceptualization, G. M. A and D. F.; methodology, G. M. A. and D. F.; software, G. G. and G. M. A.; validation, F. M. D. P. and G. M. A.; formal analysis, D. F.; investigation, G. M. A. and G. G.; data curation, G. G.; writing---original draft preparation, D. F. and G. G.; writing---review and editing, G. M. A., F. M. D. P. and M. S.; supervision, M. S. and All authors have read and agreed to the published version of the manuscript.}

\funding{
D. F. would like to thank the funding of the
European Union-NextGenerationEU through the "Quantum
Busses for Coherent Energy Transfer" (QUBERT) project, in
the framework of Curiosity Driven 2021 initiative of the University of Genova.
F.M.D.P. was supported by
the Università degli Studi di Catania, Piano di Incentivi per la Ricerca di Ateneo 2020/2022 (progetto QUAPHENE), and
PNRR MUR project PE0000023-NQSTI.
}

\institutionalreview{Not applicable.}

\informedconsent{Not applicable.}

\dataavailability{Data are available from the authors upon request.} 

\acknowledgments{The authors thank A. Crescente, G. Falci, L. Giannelli, and E. Paladino for illuminating discussions and fruitful comments.}

\conflictsofinterest{The authors declare no conflict of interest.} 

\abbreviations{Abbreviations}{
The following abbreviations are used in this manuscript:\\
\vspace{-6pt}

\noindent 
\begin{tabular}{@{}ll}
QB & Quantum Battery\\
TLS & Two-level system\\
LMG &  Lipkin-Meshkov-Glick\\
ESQPT &  Excited State Quantum Phase Transition
\end{tabular}}

\abbreviations{Nomenclature}{
{{{List} of the Relevant Parameters in the Model:\\}}

\noindent 
\begin{tabular}{@{}ll}

\end{tabular}}

\appendixtitles{yes}
\appendixstart
\appendix

\section{Derivation of the effective LMG Hamiltonian}\label{AppA}

\subsection{Basics of the Schrieffer–Wolff transformation}

Given a generic Hamiltonian $\hat{\mathcal{H}}$ and an operator $\hat{\mathcal{S}}$ it is possible to define

\begin{equation}
    \hat{\mathcal{H}}'=e^{\hat{\mathcal{S}}}\hat{\mathcal{H}}e^{-\hat{\mathcal{S}}}. 
\end{equation}
According to the Baker-Campbell-Hausdorff relation, one can develop the exponential operators in such a way that
\begin{equation}
    \hat{\mathcal{H}}'=\hat{\mathcal{H}}+\left[\hat{\mathcal{S}},\hat{\mathcal{H}}\right]+\frac{1}{2!}\left[\hat{\mathcal{S}},\left[\hat{\mathcal{S}},\hat{\mathcal{H}}\right]\right]+\ldots.
\end{equation}
On a general ground, one can now assume that it is possible to decompose $\hat{\mathcal{H}}$ in a free and an interacting term, namely 
\begin{equation}
\hat{\mathcal{H}}=\hat{\mathcal{H}}_0+\hat{\mathcal{V}}.
\end{equation}
By properly choosing the operator $\hat{\mathcal{S}}$ in such a way to satisfy
\begin{equation}
\left[\hat{\mathcal{H}}_0,\hat{\mathcal{S}}\right]=\hat{\mathcal{V}}
\end{equation}
one obtains
\begin{equation}
    \hat{\mathcal{H}}'=\hat{\mathcal{H}}_0+\frac{1}{2}\left[\hat{\mathcal{S}},\hat{\mathcal{V}}\right]+\frac{1}{2}\left[\hat{\mathcal{S}},\left[\hat{\mathcal{S}},\hat{\mathcal{V}}\right]\right]+\ldots.
\end{equation}
Assuming now that both $\hat{\mathcal{S}}$ and $\hat{\mathcal{V}}$ depend linearly on a given coupling constant one can consider a second order expansion leading to an effective Hamiltonian of the form 
\begin{equation}
    \hat{\mathcal{H}}_{\rm eff}=\hat{\mathcal{H}}_0+\frac{1}{2}\left[\hat{\mathcal{S}},\hat{\mathcal{V}}\right].
\end{equation}
In the following, we will apply this general scheme to the Dicke model assuming both single- and two-photon interaction. 

\subsection{Dicke model with single-photon coupling}
In the case of the Dicke model with conventional dipole matter-radiation coupling~\cite{Dicke54, Schleich_Book} in Eq.~(\ref{Dicke}) one can identify (using the notation of the main text)
\begin{eqnarray}
\hat{\mathcal{H}}_{0}&=&\omega_z\hat{S}_z+\omega_c\hat{a}^{\dagger}\hat{a},\\
\hat{\mathcal{V}}&=& 2\lambda\hat{S}_x\left(\hat{a}^{\dagger}+\hat{a}\right),
\label{V}
\end{eqnarray}
and consequently
\begin{equation}
\hat{\mathcal{S}}=\frac{\lambda}{\omega_c-\omega_z}\left[\hat{S}_-\hat{a}^{\dagger}-\hat{S}_+\hat{a}\right]+\frac{\lambda}{\omega_c+\omega_z}\left[\hat{S}_+\hat{a}^{\dagger}-\hat{S}_-\hat{a}\right],
\label{S}
\end{equation}
with
\begin{equation}
    \hat{S}_{\pm}=\hat{S}_{x}\pm i \hat{S}_y.
\end{equation}
Notice that both Eq.~(\ref{V}) and~(\ref{S}) are linear in the matter-radiation coupling $\lambda$ as required by the previous discussion.  
Using the Schrieffer–Wolff transformation discussed above one obtains the effective Hamiltonian

\begin{equation}
    \hat{\mathcal{H}}_{\rm eff}=\omega_z \hat{S}_z+\omega_c \hat{a}^{\dagger}\hat{a}-\frac{2\lambda^2\omega_z}{\omega_c^2-\omega_z^2}\hat{S}_z\left(\hat{a}^{\dagger}+\hat{a}\right)^2-\frac{4\lambda^2\omega_c}{\omega_c^2-\omega_z^2}\hat{S}_x^2,
\end{equation}
which is a good approximation of the Dicke Hamiltonian in the so called  dispersive regime $\lambda\ll|\omega_{c}-\omega_{z}|$~\cite{Schleich_Book}. Further considering the limit $\omega_{c}\gg \omega_{z}$ one finally has 
\begin{equation}
    \hat{\mathcal{H}'}_{\rm eff}=\omega_z \hat{S}_z+\omega_c \hat{a}^{\dagger}\hat{a}-\frac{4\lambda^2}{\omega_c}\hat{S}_x^2.
    \label{H_1ph_prime}
\end{equation}
Getting now rid of the cavity contribution it is possible to map Eq.~(\ref{H_1ph_prime}) into Eq.~(\ref{H}) by defining 
\begin{equation}
g=\frac{4\lambda^2}{\omega_c}. 
\end{equation}

\subsection{Dicke model with two-photon coupling}

This more exotic interaction can be obtained by properly engineering both atomic and solid state systems~\cite{Felicetti15, Felicetti18}. Here, one can write the Hamiltonian 
\begin{equation}
\hat{H}^{2ph}_{\rm {Dicke}}=\omega_z\hat{S}_z+\omega_c\hat{a}^{\dagger}\hat{a}+ 2\lambda\hat{S}_x\left[\left(\hat{a}^{\dagger}\right)^{2}+\left(\hat{a}\right)^{2}\right],
\label{Dicke_2ph}
\end{equation}

identifying 
\begin{eqnarray}
\hat{\mathcal{H}}_{0}&=&\omega_z\hat{S}_z+\frac{\omega_c}{2}\hat{a}^{\dagger}\hat{a},\\
\hat{\mathcal{V}}&=&2\lambda\hat{S}_x\left[\left(\hat{a}^{\dagger}\right)^2+\hat{a}^2\right],
\end{eqnarray}
and consequently
\begin{equation}
\hat{\mathcal{S}}=\frac{\lambda}{\omega_c-\omega_z}\left[\hat{S}_-\left(\hat{a}^{\dagger}\right)^2-\hat{S}_+\hat{a}^2\right]+\frac{\lambda}{\omega_c+\omega_z}\left[\hat{S}_+\left(\hat{a}^{\dagger}\right)^2-\hat{S}_-\hat{a}^2\right]. 
\end{equation}

According to the previous analysis, one obtains the effective Hamiltonian
\begin{equation}
    \hat{\mathcal{H}}_{\rm eff}^{2ph}=\omega_z \hat{S}_z+\frac{\omega_c}{2} \hat{a}^{\dagger}\hat{a}-\frac{2\lambda^2\omega_z}{\omega_c^2-\omega_z^2}\hat{S}_z\left[\hat{a}^2+\left(\hat{a}^{\dagger}\right)^2\right]^2-\frac{4\lambda^2\omega_c}{\omega_c^2-\omega_z^2}\hat{S}_x^2\left(\hat{a}\hat{a}^{\dagger}+\hat{a}^{\dagger}\hat{a}\right),
\end{equation}
which again represents a good approximation of the two-photon Dicke Hamiltonian in the dispersive limit $\lambda\ll|\omega_{c}-\omega_{z}|$.

Further considering the limit $\omega_{c}\gg \omega_{z}$ one finally has 
\begin{equation}
    \hat{\mathcal{H}'}_{\rm eff}^{2ph}=\omega_z \hat{S}_z+\frac{\omega_c}{2} \hat{a}^{\dagger}\hat{a}-\frac{4\lambda^2}{\omega_c}(2\hat{a}^{\dagger}\hat{a}+1)\hat{S}_x^2.
    \label{H_2ph_prime}
\end{equation}

For proper initial states of the photon in the cavity, it is possible to map Eq.~(\ref{H_2ph_prime}) into Eq.~(\ref{H}) (up to a constant) by defining 
\begin{equation}
g=\frac{4\lambda^2}{\omega_c}(2 n+1),  
\end{equation}
with $n$ the averaged number of photons into the cavity. Notice that in this case the renormalization of the effective coupling due to the presence of the cavity photons can be exploited to increase the value of $g$ with respect to the previous case.

\section{Perturbative approach to the weak coupling case}
\label{AppB}

It is possible to study the dynamics induced by Eq.~(\ref{H}) and the figures of merits discussed in the main text by applying the time-dependent perturbation theory in terms of the Dyson series.

Working in the Interaction picture (indicated in the following by the apex $I$), the time evolution operator $\hat{U}_I(t,0)$ from the initial time $t=0$ to a generic time $t$ satisfies the differential equation 
\begin{equation}
    i \partial_t\hat{U}_I(t,0)=\hat{H}^{I}_{\rm int}(t)\hat{U}_I(t,0),
\end{equation}
with
\begin{eqnarray}
\hat{H}^{I}_{\rm int}(t)&=&-g e^{ i\omega_zt\hat{S}_z}\hat{S}_x^2e^{- i\omega_zt\hat{S}_z} \nonumber\\
&=&-g e^{ i\omega_zt\hat{S}_z}\hat{S}_xe^{- i\omega_zt\hat{S}_z}e^{ i\omega_zt\hat{S}_z}\hat{S}_xe^{- i\omega_zt\hat{S}_z} \nonumber \\
&=&-g\left[\cos{\left(\omega_zt\right)}\hat{S}_x^2-\sin{\left(\omega_zt\right)}\cos{\left(\omega_zt\right)}\left(\hat{S}_x\hat{S}_y+\hat{S}_y\hat{S}_x\right)+\sin{\left(\omega_zt\right)}\hat{S}_y^2\right].
\end{eqnarray}

Imposing the initial condition $\hat{U}_I(0,0)=1$ and formally integrating the above equation we have
\begin{equation}
    \hat{U}_I(t,0)=1-i\int_0^t\hat{H}^{I}_{\rm int}(t')\hat{U}_I(t',0)dt'.
\end{equation}
We can solve iteratively this equation in terms of Dyson's series~\cite{Sakurai_Book}
\begin{equation}
    \hat{U}_I(t,0)=1-i\int_0^t\hat{H}^{I}_{\rm int}(t')dt'+\left(-i\right)^2\int_0^tdt'\int_0^{t'}dt''\hat{H}^{I}_{\rm int}(t')\hat{H}^{I}_{\rm int}(t'')+\ldots.
\end{equation}

In this picture, the energy stored in the QB can be written as
\begin{eqnarray}
    E_{B}(t)&=&\langle\frac{N}{2},-\frac{N}{2}|e^{-i\hat{H}_0t}\hat{U}_I^{\dagger}(t,0)\hat{H}_B\hat{U}_I(t,0)e^{i\hat{H}_0t}|\frac{N}{2}, -\frac{N}{2}\rangle \nonumber \\
    &-&\langle\frac{N}{2}, -\frac{N}{2}|\hat{H}_B|\frac{N}{2}, -\frac{N}{2}\rangle  \nonumber \\
    &=&\langle\frac{N}{2},-\frac{N}{2}|\hat{U}_I^{\dagger}(t,0)\hat{H}_B\hat{U}_I(t,0)|\frac{N}{2}, -\frac{N}{2}\rangle+\frac{\omega_{z}}{2} N.
\end{eqnarray}
After some long and tedious algebra, one finally obtains
\begin{equation}
    E_{B}(t)=\frac{g^2}{8}\frac{1-\cos{2\omega_z t}}{\omega_z}\left(N^2-N\right),
\end{equation}
and consequently 
\begin{equation}
    P(t)=\frac{E(t)}{t}=\frac{g^2}{8}\frac{1-\cos{2\omega_z t}}{t\omega_z}\left(N^2-N\right).
\end{equation}

\section{Comments on the classical limit} \label{AppC}
In the main text, we have considered the scaling at a finite number of TLSs, without addressing the thermodynamical limit needed for a proper study of the ESQPT. Some hints in this direction can be derived starting from the classical version of the Hamiltonian in Eq.~(\ref{H}). This can be obtained replacing the operators $\hat{S}_{x,y,z}$ with the classical angular momentum components
\begin{eqnarray}
S_{x}&=& \frac{N}{2}\sin\theta \cos\varphi,\\
S_{y}&=& \frac{N}{2}\sin\theta \sin\varphi,\\
S_{z}&=&\frac{N}{2}\cos\theta.
\end{eqnarray}
This leads to the classical Hamiltonian 
\begin{equation}
H_{\rm cl}=\omega_{z} \frac{N}{2} \cos \theta- g \frac{N^{2}}{4} \sin^2\theta \cos^{2}
\varphi.
\end{equation}
One can identify now two conjugated variables~\cite{Andolina19} 
\begin{equation}
q=N \cos\theta,
\end{equation}
\begin{equation}
 p=\varphi,
\end{equation}
in such a way that 
\begin{equation}
H_{\rm cl}=\frac{\omega_{z}}{2}q-\frac{g}{4}N^{2} \cos^{2}p + \frac{g}{4}q^{2}\cos^{2}p.
\end{equation}

Recalling the equations of motion
\begin{eqnarray}
\frac{dq}{dt}&=&\frac{\partial H_{\rm cl}}{\partial p}=\frac{g}{4}N^{2}
\sin{2p}-\frac{g}{4}q^{2} \sin{2 p},\\
\frac{dp}{dt}&=&-\frac{\partial H_{\rm cl}}{\partial q}=-\frac{\omega_{z}}{2}+\frac{g}{2}q \cos^{2}{p}.
\end{eqnarray}

Further manipulating the above Equations introducing rescaled and shifted variables
\begin{equation}
q=N\tilde{q},
\end{equation}
\begin{equation}
p=\tilde{p}-\frac{\omega_{z}}{2}t,
\end{equation}
one has
\begin{eqnarray}
\frac{d \tilde{q}}{dt}&=& \frac{g}{4}N (1-\tilde{q}^{2})\sin\left( 2 \tilde{p}-\omega_{z}t\right),\\
\frac{d \tilde{p}}{dt}&=&-\frac{g}{2}
N \tilde{q}\cos^{2}\left(\tilde{p}-\frac{\omega_{z}}{2}t \right),
\end{eqnarray}
where clearly emerge the dependence on the renormalized interaction parameter 
\begin{equation}
G=g N
\end{equation}
which plays a major role in the discussion about the universality reported in the main text.

\begin{adjustwidth}{-\extralength}{0cm}

\reftitle{References}

\end{adjustwidth}
\end{document}